\RequirePackage{fix-cm}

\documentclass[smallextended]{svjour3} 
\usepackage{amsmath, amssymb} 
\usepackage{hyperref} 
\smartqed  

\begin{document}

\title{Editorial note to: \\ Erwin Schr\"odinger, Dirac electron in the gravitational field I}


\author{Bernard~S.~Kay}


\institute{Department of Mathematics, University of York, York YO10 5DD
             \email{bernard.kay@york.ac.uk}         
} 

\date{Received: date / Accepted: date}

\maketitle

\begin{abstract}
We aim to give a mathematical and historical introduction to the 1932
paper ``Dirac equation in the gravitational field I'' by Erwin
Schr\"odinger on the generalization of the Dirac equation to a curved
spacetime and also to discuss the influence this paper had on
subsequent work. The paper is of interest as the first place that the
well-known formula $g^{\mu\nu}\nabla_\mu\nabla_\nu +m^2 + R/4$ was
obtained for the `square' of the Dirac operator in curved
spacetime. This formula is known by a number of names and we explain
why we favour the name `Schr\"odinger-Lichnerowicz formula'. We also
aim to explain how the modern notion of `spin connection' emerged from
a debate in the physics journals in the period 1929-1933. We discuss
the key contributions of Weyl, Fock and Cartan and explain how and why
they were partly in conflict with the approaches of Schr\"odinger and
several other authors. We reference and comment on some previous
historical accounts of this topic.
\end{abstract}

\keywords{Schr\"odinger \and history \and Dirac equation \and spin connection \and curved spacetime \and Schr\"odinger-Lichnerowicz formula \and square of Dirac operator}

\section{Introduction}
\label{intro}
This Golden Oldie is a translation from the original German of \cite{Schr32}.  
(Despite the numeral `I' in the paper's title, there doesn't appear to have been a second instalment.)   As we shall see, this is an, in some ways flawed, but, nevertheless both mathematically and historically interesting, paper.  It seems to have been the first place that the formula (Equation \eqref{Diracsquare} below) was obtained for the square of the curved-spacetime Dirac operator. It is also one of a number of papers by a number of authors from the period 1928-33 which embodied an interesting scientific debate and from which the important notion of \textit{spin connection} emerged.   It is \textit{not} one of those Golden Oldies about which one could justifiably say (with an, in consequence, rather short editorial note):  ``If you want to learn this topic, then read this original paper.'' (And that can arguably also not be said of any other of those several 1928-1933 papers although, as we shall relate, two of the 1929 papers do give a correct account of the topic.)  The reader would surely be better off with the modern textbook literature for that purpose.  But there seems to be more worth saying than has been said in previous accounts of the history of the notion of spin connection, and it is hoped that what follows will go some way towards filling the gap.  

\medskip

\noindent
{\bf Added Note October 2022:} Since posting v6 of this paper on the arXiv and publishing it as \cite{KayGRG} it was pointed out to me by Gary Gibbons and (indirectly) by Mike Stone that aside from the rediscovery of Schr\"odinger's formula for the `square' of the curved-spacetime Dirac Operator in 1962/3 by Andr\'e Lichnerowicz, it was also rediscovered in 1963 by Asher Peres.   Gary Gibbons and I have prepared a brief addendum to the present paper which mainly discusses that latter rediscovery and which it is intended to publish separately.   The present v7 is identical to v6 except that the essential content of that addendum is appended before the bibliography.   The five papers newly cited in the addendum have also been added at the end of the bibliography.

\section{Background}
\label{sec1}
The proposal of matrix mechanics by Heisenberg in 1925, together with that of wave mechanics by Schr\"odinger in 1926, surely constitute one of the biggest ever upheavals in our understanding of the laws of physics.  Its/their potential implications for physics, and also for chemistry and mathematics, were vast.  

The opportunities these proposals opened up were very apparent to physicists and mathematicians at the time and it's remarkable how rapid was the progress in the next few years.
The Hydrogen atom was quickly solved (by Pauli in 1926) and the basis for quantum chemistry laid, but also, around 1926-30 the basis was laid for quantum field theory (see \cite[Chapter 1]{Weinberg}).   On the side of the mathematical formalism, Dirac (after an earlier attempt by Schr\"odinger himself) showed the equivalence of Heisenberg's and Schr\"odinger's theories in 1927, later (1932) to be re-worked in a fully mathematically rigorous way by von Neumann.   

And, in 1928, Dirac came up with his celebrated equation
\begin{equation}
\label{Dirac}
i\gamma^a\partial_a\psi - m\psi=0,
\end{equation}
where
\begin{equation}
\label{gammaflat}
\{\gamma^a, \gamma^b\} = 2\eta^{ab},
\end{equation}
where $\eta_{ab}$ is the Minkowskian metric (which we shall take here to have signature 
$(+, -,-,-)$).
Dirac's equation, by itself, opened up a vast range of further questions.   Amongst these,  an interesting and important matter of principle was raised by the

\medskip

\noindent
{\bf Question} \textit{How, even just locally, can the notion of a Dirac wavefunction $\psi$ be generalized to a general curved spacetime?  And, assuming that is solved, how (again, even just locally) can one generalize Dirac's equation and thus achieve compatibility with Einstein's theory of general relativity (which dated back to 1915)?} 

\medskip

The first part of our Question, in modern language, amounts to the problem of equipping our spacetime with a \textit{spin structure} and, if we are only interested in doing things locally, presents no difficulty.  But the second requires the generalization of the `$\partial_a$' of \eqref{Dirac} to a notion of covariant derivative suitable to act on the spinor $\psi$ -- or, in other words, the provision of a suitable notion of parallel-transport for $\psi$, or, in yet other words, the provision of a suitable notion of `spin connection'. 

Of course, with the mathematical concepts that have been in our possession since 
the 1950s\footnote{Here is some basic information on the history of Riemannian geometry, fibre bundles, gauge theories and spinors since 1950:  The reformulation of earlier notions of \textit{connection} in terms of a covariant derivative operator (say in the direction of a vector $X$) $\nabla_X$ ($=X^a\nabla_a$) satisfying linearity, additivity, Leibniz rule etc., as explained in most modern general relativity textbooks, is based on the work of Koszul in 1950.    Also in 1950, Ehresmann (a student of Cartan) reworked and generalized to what we would call general gauge (Lie) groups, the notion of connection in terms of a notion of horizontality in a principal fibre bundle -- see e.g. \cite{BishCritt} or \cite{Trautman} and, for a brief account of how the history of gauge theories slots in to the history of the theory of fibre bundles, see Section I in \cite{Neeman}.   
The notion of spin structure (for $SO(n)$) was introduced by Haefliger (a student of Ehresmann) in 1956 who also found a criterion (the vanishing of the second Stiefel-Whitney class) for its global existence on an orientable Riemannian manifold.   For the history of spinors, see \cite[Section 6b]{Berger} and/or watch the youtube video \cite{Atiyah}.}
this is no longer a challenge:  In one way of saying things,
one can start with the usual (torsion-free) metric connection of general relativity, say viewed as a notion of horizontality on the bundle, $B$, of Lorentz frames of our spacetime (whose structure group is the Lorentz group).  This will induce a connection on the (always at least locally defined) bundle of spin frames (which doubly covers $B$ and has, as structure group the double covering, $SL(2,\mathbb{C})$, of the Lorentz group) and that, in turn, will, by a standard construction, induce a connection on the associated bundle of spinors (which again always exists at least locally) in which our Dirac field, $\psi$, is a cross-section.  

In terms of a vierbein, $e^a_{\ \mu}$ (which coordinatises $B$ and which satisfies $e^a_{\ \mu}(x)e^b_{\ \nu}(x)\eta_{ab} = g_{\mu\nu}$, $e^\mu_{\ a}(x)e^\nu_{\ b}(x)g_{\mu\nu}=\eta_{ab}$; and we note that we shall raise and lower Latin indices with $\eta_{ab}$, and Greek indices with $g_{\mu\nu}$) the resulting connection, when re-expressed as a covariant derivative operator, may be written (see e.g.\ \cite{BirrellDavies,Lawrie,GSW}):
\begin{equation}
\label{covderiv}
\nabla_\mu\psi = \partial_\mu\psi + \Gamma_\mu\psi
\end{equation}
where
\begin{equation}
\label{Gammamu}
\Gamma_\mu(x) = -\frac{i}{4}\omega_{ab\mu}(x)\sigma^{ab}
\end{equation}
where $\omega_{ab\mu}$ are the components of the spin connection, given in terms of 
the usual Christoffel symbol for the (torsion free) metric connection, $\Gamma^\nu_{\rho\mu}$, by
\begin{equation}
\label{omegaabmu}
\omega^a_{\ b\mu}=e^a_{\ \nu}\partial e^\nu_{\ b}/\partial x^\mu + e^a_{\ \nu} e^\rho_{\ b}\Gamma^\nu_{\rho\mu}.
\end{equation}
Here $\sigma^{ab}$ (defined so that, if $\lambda^a_{\ b}$ is the generator of Lorentz transformation on 4-vectors, then $-\frac{i}{4}\lambda_{ab}\sigma^{ab}$ is the generator of Lorentz transformations on Dirac wavefunctions) is given by
\begin{equation}
\label{sigmaab}
\sigma^{ab} = \frac{i}{2}[\gamma^a, \gamma^b].
\end{equation}

The answer to the second part of our Question is then that the Dirac equation should be replaced, in a curved spacetime, by the equation
\begin{equation}
\label{DiracGR}
i\gamma^\mu\nabla_\mu\psi - m\psi=0.
\end{equation}
where $\nabla_\mu$ is as in \eqref{covderiv} and 
\begin{equation}
\label{gammamu}
\gamma^\mu=e^\mu_{\ a}\gamma^a.
\end{equation}
We note, in passing, that, in view of \eqref{gammaflat}, the $\gamma^\mu$ satisfy
\begin{equation}
\label{gammacurved}
\{\gamma^\mu, \gamma^\nu\} = 2g^{\mu\nu}.
\end{equation}

Let us also note, for later reference, that the familiar equation 
\[
[\nabla_\alpha, \nabla_\beta]v_\gamma = R_{\alpha\beta\gamma}^{\ \ \ \ \delta} v_\delta
\]
for the commutator of two covariant derivatives on (say) a covariant vector field, $v_\alpha$, has, as an analogue, the equation
\[
[\nabla_\alpha, \nabla_\beta]\psi = -\frac{i}{4}R_{\alpha\beta cd}\sigma^{cd}\psi
\]
for a spinor field, $\psi$, and this formula can be inferred just from the fact that $\nabla_\alpha$ has the general properties of a connection without any need to invoke an explicit formula for $\Gamma_\mu$.    Moreover, by \eqref{sigmaab} and \eqref{gammamu}, this may be rewitten as
\begin{equation}
\label{Rgammagamma}
\nabla_{[\alpha}\nabla_{\beta ]}\psi = \frac{1}{8}R_{\alpha\beta\delta\eta}\gamma^\delta \gamma^\eta\psi
\end{equation}
which, we notice, no longer makes any reference to a choice of vierbein.

But back in 1928, all this was yet to come.   And, even though the theory of spinors on vector spaces had been developed considerably by Cartan (who, already in 1913, had discovered what would later be understood as spinor representations of Lie Algebras) and by van der Waerden (see e.g.\ his article `Spinoranalyse' \cite{vdWaerden29} and, for a historical account,  \cite{Schneider}) and others, the answer to our above Question posed a serious challenge.    The answer emerged from a debate in the physics journals and is fascinating to read, involving a number of wrong turnings and unnecessary detours as well as misunderstandings and disagreements. (Other historical accounts can be found in \cite[Section 12]{vdWaerden60}, \cite{Kichenassamy}, \cite{Schneider},
and \cite[Section 7.2.2]{Goenner} some of which, however, appear to differ from us here in their detailed conclusions\footnote{While, in \cite[Section 12]{vdWaerden60}, van der Waerden (fairly) points out that the paper \cite{InfeldvdW} by Infeld and van der Waerden follows and further simplifies Weyl \cite{WeylNAS,WeylZ} and Fock \cite{Fock}, and that the approaches of Tetrode \cite{Tetrode} and of Schr\"odinger \cite{Schr32} and of Bargmann \cite{Bargmann} are more complicated, we caution the reader that (despite what Weyl and Cartan had taught us much earlier! -- see below) it misleadingly says that the approaches of Weyl and of Fock, and of Tetrode and of Schr\"odinger and of Bargmann are all ``\textit{equivalent from the physical point of view}''. We urge similar caution regarding similar misleading statements in other historical surveys -- e.g.\ in \cite{Kichenassamy} and \cite{Schneider}.

For more about Schr\"odinger's general motivations and interests in the relevant period, which mention, but are not confined to, his 1932 paper, see \cite{Ruger}, \cite{Urbantke} and \cite{Dorling}.  And for more about the general motivations and interests of theoretical physicists of that period, we recommend the blog article \cite{Motl} by Lubos Motl.} on the matter of Schr\"odinger's paper and the several papers that it refers to.)  And, though it is now clear that the answer (i.e.\ essentially the content of Equations \eqref{covderiv}, \eqref{Gammamu}, \eqref{omegaabmu}, \eqref{sigmaab} above) had appeared (in the papers of Weyl \cite{WeylNAS,WeylZ} and Fock \cite{Fock}) in 1929, it seems to have taken quite a few years before that answer came to be generally accepted as the standard answer in the physics literature (and for less satisfactory viewpoints such as that of \cite{Schr32} to fall out of the mainstream -- see e.g.\ where we mention the work of Brill and Wheeler below).   It took a similarly long time before that answer came to be incorporated, as an accepted part of the general folklore, in mathematics (see e.g.\ \cite[Section 6b]{Berger} for the history and \cite{LawMich} for a modern (1989) textbook treatment).
 
Once the solution had been digested though, the opportunities it opened up were enormous and included, in 1963, the important (Riemannian) Dirac-operator example of the Atiyah-Singer index theorem \cite{Atiyah63}, followed by Lichnerowicz' proof \cite{Lichner63}  (more about which below) of the vanishing of the index (i.e.\ the Hirzebruch $\hat A$-genus) in that example for even-dimensional compact Riemannian manifolds admitting spin structures with non-negative (but not identically zero) scalar curvature --  which (together with the supergravity-based arguments for positive mass due to Deser and Teitelboim and Grisaru [as cited in \cite{Witten}]) was one of the acknowledged influences on Witten's 1981 alternative proof \cite{Witten} of Schoen and Yau's positive mass theorem.    Plus Connes' noncommutative geometry \cite{Connes} and (via \cite{InfeldvdW}) the applications of spinors to General Relativity due to Newman, Penrose and others \cite{PenRind}, \cite[Chapter 13]{Wald} $\dots$, not to mention the obvious relevance of the curved spacetime Dirac equation to quantum theory in curved spacetime (on which, by the way, Schr\"odinger also wrote a pioneering paper \cite{Schr39} in 1939) and the r\^ole the Dirac equation in curved spacetime plays in supersymmetry, supergravity (see also Footnote \ref{SG}), string theory, and much more.

\section{Tetrode and Wigner, Weyl and Fock, Schr\"odinger, and Cartan}

We now turn to a detailed discussion of the Schr\"odinger paper.  We will attempt to indicate what it did that was original and had influence on later work and/but also to indicate its shortcomings.  To this end, it is useful to begin with a critical examination of what Schr\"odinger writes in his introduction: 

He begins by citing a number of papers by previous authors -- Wigner, Tetrode, Fock, Weyl, Zaycoff and Podolsky.\footnote{A number of these papers -- namely, those \cite{Tetrode,Fock,WeylZ} of Tetrode, Fock and Weyl, as well as that \cite{Schr32} of Schr\"odinger and that \cite{InfeldvdW} of Infeld and van der Waerden -- are reprinted in the recent volume \cite{Half} along with some commentary on (\textit{inter alia}) the work of these authors in the article by Alexander Blum which forms Chapter 5 of that volume.  I thank Alexander Blum for drawing my attention to this reference and also for helpful conversations about the work of Tetrode and Wigner.}  (In our language) he points out that most of them use vierbeins and remarks -- seemingly of all the papers that use vierbeins -- that ``\textit{it is a little bit difficult to recognize whether Einstein's idea of teleparallelism, to which reference is partly made, really enters or whether one is independent of it}'' and he explains that anyway he prefers not to use vierbeins because they are ``\textit{more complicated}'' (i.e.\ than tensors).   On the other hand, he commends one of the authors, namely Tetrode, for replacing the commutation relations \eqref{gammaflat} by the curved spacetime form \eqref{gammacurved}  without any mention of vierbeins.    It helps here, first, to briefly describe what Tetrode, and, after him, Wigner do:  Basically, in \cite{Tetrode}, Tetrode writes down Dirac's equation in a curved spacetime as in \eqref{DiracGR}, replacing the $\gamma^a$ of \eqref{gammaflat} by the $\gamma^\mu$ of \eqref{gammacurved}.   (Incidentally, the fact that the relations \eqref{gammacurved} are related to those of \eqref{gammaflat} by \eqref{gammamu} seems to be the essential content of the paper, \cite{FockIvanenko}, of Fock and Ivanenko.)  However, while Tetrode replaces the Minkowskian partial derivative $\partial/\partial x^a$ by a general coordinate partial derivative $\partial/\partial x^\mu$, he doesn't attempt to replace that with a covariant derivative of any sort.  He simply notes that it seems difficult to see what can replace the invariance property of the flat spacetime Dirac equation under Lorentz transformations in the curved spacetime case. Then, in \cite{Wigner},  Wigner argues that this difficulty could be resolved if one were to relate Tetrode's curved spacetime gamma matrices to Dirac's original gamma matrices using a `vierbein' (as in \eqref{gammamu}) -- however not the usual notion of vierbein, but rather the notion of `vierbein' used in Einstein's teleparallelism theory\footnote{Ironically, teleparallelism currently seems to be in the midst of one of its periodic revivals in work on alternative theories of gravity.  For its early history, see e.g.\ \cite[Sections 6.4 and 7.2]{Goenner}}.  Whereas with the usual notion of vierbein we consider in general relativity, the Lorentz group acts as a local gauge group, in the Einstein teleparallel theory it consists of four globally defined vector fields, which transform under a single global action of the Lorentz group.   Wigner shows that a suitably symmetrized version of Tetrode's Dirac equation (where $\partial/\partial x^\mu$ is replaced by a `covariant derivative' which, however, acts on spinors as if they were scalars) is invariant under this single global action of the Lorentz group.

Schr\"odinger (rightly in the modern view) doesn't want to adopt such a theory, based on teleparallelism.  But he errs in seeming to imply that all the papers which use vierbeins rely on teleparallelism.  Fock and Weyl don't!\footnote{Podolsky \cite{Podolsky} and Zaycoff \cite{Zaycoff} both appear to be concerned with building/commenting on what Fock and Weyl had already done and so we may omit further discussion of them.}    In April 1929, Weyl gives in \cite{WeylNAS} (see also the slightly later, fuller version \cite{WeylZ}) what we would regard as the right solution to our Question for his two-component version of the massless Dirac equation (which is introduced in the same work) while in July 1929, Fock gives essentially the same solution for the original version \eqref{Dirac} of the massive Dirac equation.  (For a historical discussion of the work of Weyl and of Fock, see \cite{Scholz}.)   Furthermore, regarding Schr\"odinger's wish to anyway avoid vierbeins because they are complicated, well it is clear to us now that one can't avoid them!  At least not if one wants to have an explicit expression for the covariant derivative of a spinor wave function.      

Indeed anything that pertains to the bundle of spinors (including the question of the global existence of Dirac wavefunctions themselves) requires the use of vierbeins in the sense that it requires reference to a double covering of the bundle of Lorentz frames and, because the double cover needs to be taken, we cannot revert to the bundle of general linear frames.   

This is clear from the well known fact (\cite[Sections 85 and 177]{Cartan66} and e.g.\ \cite[Page 272]{GSW}) that, for $n>2$, (the connected component of the identity of) $GL(n,\mathbb{R})$ (even though doubly connected) has no finite-dimensional multivalued (i.e.\ `spinor') representations.  (The proof is straightforward once one observes that one can replace $GL(n,\mathbb{R})$ by $SL(n, \mathbb{R})$ and that, while this is doubly connected, its complexification, $SL(n, \mathbb{C})$, is simply connected.) 

In fact, this seems to have already been clear to Weyl back in 1929 (and maybe also to Fock) and, at least by 1937, to Cartan:   Weyl writes, in \cite{WeylNAS},

\medskip

\noindent
``\textit{We need such local cartesian axes e(a) in each point $P$ in order to be able to describe the quantity $\psi$ by means of its components $\psi_1^+$, $\psi_2^+$; $\psi_1^-$, $\psi_2^-$ , for the law of transformation of the components $\psi$ can only be given for orthogonal transformations as it corresponds to a representation of the orthogonal group which cannot be extended to the group of all linear transformations. The tensor calculus is consequently an unusable instrument for considerations involving the $\psi$.}''

\smallskip

Weyl then appends an endnote to this, writing:  ``\textit{Attempts to employ only the tensor calculus have been made by Tetrode (Z. Physik, 50, 336 (1928)); J. M. Whittaker (Proc. Camb. Phil. Soc., 25, 501 (1928)), and others; I consider them misleading.}''

In Sections 85 and 177 of his 1937 book \cite{Cartan66} `The Theory of Spinors', Cartan proves the `well known' result we mentioned above and then (without referring to Weyl) makes precise Weyl's sentence ``\textit{The tensor calculus is consequently an unusable instrument for considerations involving the $\psi$.}'' (equivalently my own above sentence
``\textit{$\dots$ we cannot revert to the bundle of general linear frames}'') with the last theorem of his book which is worth quoting in full:

\medskip

\noindent
{\bf Cartan's Theorem.} \quad \textit{With the geometric sense we have given to the word ``spinor'' it is
impossible to introduce fields of spinors into the classical Riemannian
technique; that is, having chosen an arbitrary system of co-ordinates $x^i$
for the space, it is impossible to represent a spinor by any finite number
$N$ whatsoever, of components $u_\alpha$ such that the $u_\alpha$ have covariant derivatives
of the form
\[
u_{\alpha, i}=\frac{\partial u_\alpha}{\partial x^i} + \Lambda^\beta_{\alpha i}u_\beta
\]
where the $\Lambda^\beta_{\alpha i}$ are determinate functions of $x^h$.}\footnote{\label{Pitts} We note here that it is possible, though, to consider spinors as transforming under such a formula if one allows spinors to transform at each point under a \textit{nonlinear realization} of $GL(n,\mathbb{R})$.   See \cite{OgPon,Pitts}.  I thank Brian Pitts for drawing these references to my attention.}

\medskip

\noindent
Cartan makes his reasons for stating this theorem clear on the previous page (Page 150)  where, referring specifically to Schr\"odinger's paper \cite{Schr32} in a footnote\footnote{Note that a seeming reference to the same paper of Schr\"odinger in an earlier footnote on the same page (Page 150) of \cite{Cartan66} seems to be a typographical error and to have been meant to refer to the paper of Fock; in fact the journal, volume and page numbers are those of Fock's paper.}), he writes:

\medskip

``\textit{Other physicists, not wishing to employ local Galilean reference frames, have
sought to generalize Dirac's equations by using the classical technique of
Riemannian geometry. $\dots$   We shall see that if we adopt this point of view and wish to continue to regard spinors as well-defined geometric entities, which behave as tensors in the most general sense of that term, then the generalization of Dirac's equations will become impossible.}'' 

\section{What Schr\"odinger Does}

We next discuss how Schr\"odinger manages to partly get around these objections.   (But of course there will be questions that he is unable to address.)   To attempt to avoid using vierbeins, Schr\"odinger essentially focuses, not on the covariant derivative which acts on spinor wavefunctions, but rather on the covariant derivative which acts on the gamma matrices.   In modern language, this may be understood in terms of a connection on the Clifford bundle whose fibres above each spacetime point are generated by the gamma matrices at that point.   He is right that this connection does not (need to) involve vierbeins.   This is because the structure group of this Clifford bundle is obviously the Lorentz group and not its double cover, and a connection on it can therefore be regarded as associated to a connection on the bundle of Lorentz frames and this, in its turn, can of course be obtained by restriction from a connection (with structure group $GL(4, \mathbb{R})$) on the bundle of general linear frames.   Schr\"odinger (and Fock before him) essentially argue that this covariant derivative should take the form (cf.\ Equation (S8))\footnote{We use an `S' in front of an equation number to indicate that it is an equation number in \cite{Schr32}.}
\begin{equation}
\label{gammaconn}
\nabla_\rho\gamma_\nu=\frac{\partial\gamma_\nu}{\partial x^\rho}-\Gamma^{\mu}_{\nu\rho}\gamma_{\mu}
-\Gamma_\rho\gamma_\nu+\gamma_\nu\Gamma_\rho. 
\end{equation}
and should be required to vanish!   Here, $\Gamma_\mu$ is (apart from an ambiguity in its sign [which is, we might say, the cause of the vierbein trouble] and up to the addition of an undetermined term of form $B_\mu I$ where $B_\mu$ is a covariant vector field and $I$ the identity operator -- more about this term below) taken to be the same as the $\Gamma_\mu$ of Equation \eqref{Gammamu} above which gives us the covariant derivative on spinor wave functions.   

What Schr\"odinger \textit{can't} do, however, is obtain an explicit formula for $\Gamma_\mu$ because, as we explained above, that would require the use of vierbeins.  (The reader may verify, that, indeed, nowhere in the paper is there an equation with $\Gamma_\mu$ standing alone on its left side!)\footnote{Actually, related to Footnote \ref{Pitts}, if one introduces a suitable notion for a preferred matrix square root, $r_{\mu\nu}$, of $g_{\mu\nu}$ (thought of as a 4 x 4 matrix in each coordinate system) then one \text{can} find, in terms of it, a vierbein-free formula for $\Gamma_\mu$.  However, $r_{\mu\nu}$ will necessarily transform nonlinearly under general coordinate transformations.   See again \cite{OgPon,Pitts}.  I again thank Brian Pitts for pointing this out to me.}

Let us pause to make three further historical remarks here.   First, in their paper \cite{InfeldvdW},  Infeld and van der Waerden criticise Schr\"odinger's paper for never giving an explicit formula for $\Gamma_\mu$, albeit they fail to make the (stronger) point that this would be impossible in principle without using vierbeins (which they themselves do, though, use).   In fact essentially what \cite{InfeldvdW} does is to adapt what Fock and Weyl had done to a formalism in which the Dirac equation is viewed as a pair of coupled 2-spinor equations and this was influential in (and referenced in) the later work of Penrose and Newman and others. (See e.g.\ \cite{Penrose60,Penrose65,PenRind} and \cite[Chapter 13]{Wald}).)

Secondly, an explict formula for $\Gamma_\mu$ \textit{was} later obtained in a sort of addendum to the Schr\"odinger paper, \cite{Bargmann}, written by Bargmann who was then a pre-doctoral student with Schr\"odinger in Berlin (soon after to flee Nazi Germany to Z\"urich where he obtained his PhD under the supervision of Wentzel).   Of course, to do this, it uses vierbeins -- as it must! -- and so brings us back around a circle to what Weyl and Fock had done in the first place.   (By the way, related to our `mainstream' remark above,  Brill and Wheeler's 1957 article \cite{BrillWheel} on neutrinos in gravitational fields adopts Bargmann's Schr\"odinger-inspired way of explaining spin connections [rather than Weyl's or Fock's] giving pride of place to the above Equation \eqref{gammaconn}  [and even adopting the view of Schr\"odinger (and of Fock) that part of 
$\Gamma_\mu$ may be identified with an electromagnetic 4-potential, $A_\mu$ -- as we will discuss further below].)

Lastly, it is interesting to ask what Dirac himself knew or did about generalizing his equation to curved spacetimes.   In 1935, he wrote a paper \cite{Dirac35} generalizing the Dirac equation to de Sitter space and to anti de Sitter space and in 1936, a further paper \cite{Dirac36} generalizing the Dirac equation to  ``a four-dimensional surface of a hyperquadric in five-dimensional projective space''.   These appear to have been in line with Dirac's quest for beauty in his equations, beauty here being interpreted as a high degree of spacetime symmetry.    However, it was only in 1958, with his paper \cite{Dirac58}, that Dirac addressed the question of generalizing his equation to a general curved spacetime and cited the earlier work on this topic by (i.a.) Cartan (our reference \cite{Cartan66}), Fock \cite{Fock},  Infeld and van der Waerden \cite{InfeldvdW}, Schr\"odinger (i.e.\ our main paper of interest) \cite{Schr32}, Tetrode \cite{Tetrode} and Weyl \cite{WeylZ}  -- though without indicating e.g.\ the shortcomings that we have indicated concerning the work in \cite{Tetrode} and \cite{Schr32}.   He then proceeds to give his own solution to the problem which, as he indicates, is on similar lines to Fock's solution, except that it is couched in terms of the Dirac matrices, $\alpha^i$ and $\beta$, related to the Dirac gamma matrices (i.e.\ the $\gamma^a$ of (\ref{gammaflat}) by  $\alpha^i = \gamma^0\gamma^i$ and $\beta = \gamma^0$.  Dirac claims that his approach is ``rather more direct than Fock's'' and that it has some other advantages.

Returning to Schr\"odinger's paper, while it adopts an approach that is incapable of giving an explicit formula for $\Gamma_\mu$, this doesn't prevent it from doing a number of things which don't require vierbeins and don't require such an explicit formula.  First it looks at quantities, such as $\bar\psi\gamma^\mu\psi$, but including more general tensor quantities, which are sesquilinear in $\psi$, for which the covariant derivative obviously doesn't need to involve vierbeins; e.g.\ $\bar\psi\gamma^\mu\psi$ itself is just an ordinary vector field!   

Secondly -- and this was undoubtedly the most important new result in Schr\"odinger's paper --  it obtains (see Equation (S74)) the formula (in our conventions and setting the electromagnetic field to zero):
\begin{equation}
\label{Diracsquare}
\left(g^{\mu\nu}\nabla_\mu\nabla_\nu + m^2 + \frac{R}{4}\right)\psi=0.
\end{equation}
for the square of the Dirac equation (more precisely for the middle expression in  \eqref{nearlysquare}).\footnote{\label{sqrt} The two factors of $\sqrt{g}$ in Equation (S74) appear to be unnecessary (but harmless). (\emph{Added October 2022:} See however the last paragraph of the Addendum.) The difference in sign in front of the $\frac{1}{4}R$ term between \eqref{Diracsquare} and (S74) is due to our different signature conventions.}   

Let us remark about this, first, that thirty years later, the analogue of this formula for an even dimensional Riemannian manifold was discovered by Lichnerowicz in \cite{Lichner63} where it is used as a mathematical tool to prove his theorem which we mentioned above.   

Secondly, as we indicated above and will discuss further below, Schr\"odinger's Equation (S74) is actually a generalization of \eqref{Diracsquare} to include an external electromagnetic field (with $\nabla_\mu$ now meaning  $\partial_\mu +\Gamma_\mu - ieA_\mu$) in which the term $R/4$ above is replaced by $R/4 + \frac{1}{2}\sigma^{ab}F_{ab}$ where $F_{ab}$ is the electromagnetic field-strength tensor $\partial_aA_b-\partial_bA_a$.

Thirdly, the fact that this formula can be derived without the use of vierbeins, and without the need for an explicit formula for $\Gamma_\mu$ can be seen from Equation \eqref{Rgammagamma} which, as we remarked around that equation, can also be derived without the use of vierbeins and without the need for an explicit formula for $\Gamma_\mu$.   Indeed we could multiply the (curved spacetime) Dirac equation \eqref{DiracGR} by $(-i\gamma^\mu\nabla_\mu- m)$, thus obtaining
\begin{equation}
\label{nearlysquare}
0=(-i\gamma^\mu\nabla_\mu - m)(i\gamma^\nu\nabla_\nu\psi - m\psi)=
\gamma^\mu\gamma^\nu(\nabla_{(\mu}\nabla_{\nu)} + \nabla_{[\mu}\nabla_{\nu]} + m^2)\psi
\end{equation}
which, by \eqref{Rgammagamma} is equal to
\[
\left(g^{\mu\nu}\nabla_\mu\nabla_\nu + m^2 + \frac{1}{8}R_{\mu\nu\delta\eta}\gamma^\mu\gamma^\nu\gamma^\delta \gamma^\eta\right)\psi
\]
The last term in the brackets above is easily seen to be the same as the displayed expression after Equation (S73) and we can, at this point, join the derivation sketched in Schr\"odinger's paper to see that it is equal (with our conventions) to $R/4$.

The paper of Schr\"odinger appears to be the first place that Equation \eqref{Diracsquare} appears.   In the literature, Equation \eqref{Diracsquare} is often called the ``\textit{Lichnerowicz formula}'' (see e.g. \cite{Berger}) or sometimes (in recognition of an analogous equation involving the Hodge Laplacian for differential forms, found before the Dirac equation even existed) the ``\textit{Bochner-Lichnerowicz}'' or ``\textit{Lichnerowicz-Bochner-Weitzenb\"ock formula}'' etc.   However, some  authors  (see e.g.\ \cite{Chrysikos}) call it the ``\textit{the Schr\"odinger-Lichnerowicz formula}'' and this is surely what everyone \textit{ought} to call it!

Let us mention here that generalizations of this Schr\"odinger-Lichnerowicz formula and possible applications to elementary particle physics model building have been discussed e.g.\ in \cite{AckTolks,Tolks2001,Tolks2007} where information can also be had about more mathematically sophisticated ways of thinking about, and proving, such formulae.

Let us also mention some further aspects of Schr\"odinger's paper which are of historical and/or potential scientific interest and some further connections with later work.

First of all, let us return to the parenthetical remark we made after Equation \eqref{gammaconn} about the ambiguity in $\Gamma_\mu$ when thought of as a solution to Equation \eqref{gammaconn}.    In fact, Schr\"odinger (and Fock and Weyl before him -- see again \cite{Scholz} for a historical account -- and some others who followed them later, including Brill-Wheeler \cite{BrillWheel}) interpreted that ambiguity as allowing for an external electromagnetic 4-potential in addition to an external gravitational field, so that the full Dirac equation takes the form
\begin{equation}
\label{DiracExt}
i\gamma^\mu(\partial_\mu+\Gamma_\mu - ieA_\mu)\psi - m\psi=0.
\end{equation}  
with (if one pins down $\Gamma_\mu$ as the trace-free part of the sum $\Gamma_\mu + B_\mu I$)  $-ieA_\mu$ identified with $B_\mu$.  (And hence it was natural for Schr\"odinger to derive the electromagnetic generalization of \eqref{Diracsquare}.)   On the other hand, as Schr\"odinger indicates (see around Equations (S10) and (S15)) Equation \eqref{DiracExt} (which could of course have been arrived at quite independently of consideration of Eq (S8)/\eqref{gammaconn}) suggests that the electromagnetic potential belongs to the same family of mathematical objects as the gravitational (spin-)connection $\Gamma_\mu$ and, in this sense, Schr\"odinger (together with Fock and Weyl) can be considered to have anticipated some of the ideas later explained by Utiyama (see \cite{Utiyama} and also \cite{Trautman}) and others about the close relation between gravity and gauge theories.

Equation \eqref{gammaconn} appears to have inspired Chisholm and Farwell (see e.g. \cite{Chisholm88}, \cite{Chisholm2002}) to consider analogues of this equation involving other Clifford algebras and to base on these some ideas for elementary particle theory model-building.

Lastly, let us mention that, aside from exhibiting the close analogy between gravity and electromagnetism discussed above, another motivation for Schr\"odinger's work was to try to explain the origin of mass as a gravitational effect.\footnote{\label{SG} It is noteworthy that the idea that mass might arise as a gravitational effect was also taken in Weyl's paper \cite{WeylNAS} as his excuse for considering the massless Dirac equation (so as to eliminate the negative energy solutions).    Let us also note, although it is not strictly relevant to our evaluation of Schr\"odinger's paper, that one finds in the same paper of Weyl perhaps the first ever statement of electromagnetic gauge invariance -- i.e., in his own notation, $\psi\mapsto e^{i\lambda}\psi, \phi_p \mapsto \phi_p - \frac{\partial\lambda}{\partial x_p}$.   It is also noteworthy that Weyl revisited the coupling of the Dirac equation to gravity in a further paper \cite{Weyl1950} in 1950, where he encounters the fact that in a Lagrangian formulation in which one varies the metric and the connection independently, one obtains a generalization of General Relativity (also associated \cite{Hehl76} with Einstein, Cartan, Sciama, Kibble and several other names) involving torsion, where the torsion couples to the spin density.   This has also been influential in later work, \textit{inter alia} in Supergravity \cite{DeserZumino}.} (See \cite{Ruger} for more discussion.)  Indeed, after obtaining Equation (S8) he remarks

\medskip

\noindent
``\textit{The second term seems to me to be of considerable theoretical
interest. It is, however, too small by many, many powers of ten to be
able to {\em replace}, for example, the term on the right-hand
side. For $\mu$ is the reciprocal Compton wavelength, about
$10^{11}{\rm cm}^{-1}$. At least it seems significant that one
naturally meets in the generalized theory a term at all similar to the enigmatic
mass term.}''

\medskip

\noindent
(A reference is appended to a paper by Veblen and Hoffmann on Kaluza-Klein theory presumably just to acknowledge that they get a Klein-Gordon equation with a similar $R$ correction to a mass term from different considerations.)

For some recent ideas about mass generation that appear to be related to, or perhaps inspired by, this aspect of Schr\"odinger's paper, see e.g.\ \cite{Pollock2010}.

\section{\label{Add} Addendum}

In the above main article (now published as the editorial note \cite{KayGRG} to the recent  (re-)publication \cite{EngSchr} of an English translation [by Claus Kiefer] of the 1932 paper \cite{Schr32} of Erwin Schr\"odinger) it was mentioned that the latter paper of Schr\"odinger had been (by thirty years!) the first place where the formula $g^{\mu\nu}\nabla_\mu\nabla_\nu +m^2 + R/4$ had been obtained for the `square' $(-i\gamma^\mu\nabla_\mu - m)(i\gamma^\nu\nabla_\nu\psi - m)$ of the Dirac operator.

It was also mentioned in \cite{KayGRG} that, around 30 years later, the corresponding formula was rediscovered by Andr\'e Lichnerowicz.   (As stated in \cite{KayGRG}, this was done in the special case of zero mass and on a Riemannian manifold rather than a spacetime in \cite{Lichner63} but actually the formula for the square of the massless Dirac operator also appeared in the slightly earlier paper \cite{Lichner62} in a Lorentzian context.)  Lichnerowicz used the result in \cite{Lichner63}, to prove the vanishing of the index -- i.e.\ the Hirzebruch $\hat A$-genus -- for even-dimensional compact Riemannian manifolds admitting spin structures with non-negative (but not identically zero) scalar curvature.   (This was an early application of the Atiyah Singer index theorem, when applied to the case of Riemannian manifolds with spin structures.)  And it was argued in \cite{KayGRG} that, of the several names (including ``\textit{Bochner-Lichnerowicz} formula'', ``\textit{Lichnerowicz-Bochner-Weitzenb\"ock formula}'' etc.\ as well as ``\textit{Schr\"odinger-Lichnerowicz formula}'') that had been used by other authors, the name \textit{Schr\"odinger-Lichnerowicz formula} seemed the most appropriate.

However, it was unfortunately overlooked in \cite{KayGRG} that, also in 1963, Asher Peres rediscovered the same formula in \cite{Peres63} in aid of making the point that the `squared' Dirac equation in an external gravitational field doesn't contain a  gyro-gravitational term analogous to the gyro-magnetic term (see the top of Page 10 in \cite{KayGRG}) $\tfrac{1}{2}\sigma^{ab}F_{ab}$ in the `squared' Dirac equation in an external electromagnetic field.  Also, Peres pointed out that it is not analogous to the equation for a classical spinning particle in a curved spacetime where there is a term of form $\frac{1}{2}R_{\alpha\beta\gamma\delta}v^\beta S^{\gamma\delta}$ where $v^\alpha$ is the velocity vector and $S^{\alpha\beta}$ the angular momentum tensor of the classical spinning particle.\footnote{It is not really correct, though, to say, as Peres does that ``Dirac particles $\dots$ cannot be used to test general relativity.'' since of course, say in a coordinate system, the term $g^{\mu\nu}\nabla_\mu\nabla_\nu$ in the `squared' Dirac equation is not the same as the scalar Laplacian.}  Actually, in obtaing the formula for the `square' of the Dirac operator, \cite{Peres63} makes reference to the equations in Peres's earlier paper \cite{Peres62} which also discusses the theory associated with the names of Einstein, Cartan, Kibble, Sciama and Weyl and others (see Footnote 11 in \cite{KayGRG}) of gravity with torsion and thus, arguably, Peres deserves to have his name included in that list of names too.   (Somewhat ironically though, \cite{Peres63} does not mention that the theory involving torsion discussed in \cite{Peres62} \emph{would} lead to a gravitational counterpart of spin-orbit coupling.) 

This unfortunate omission leads us to reflect on the often noted fact that the naming of theorems in mathematics and of results in physics is often the result of historical quirks and accidents and frequently unsatisfactory or unfair in one way or another.   In the case of the formula for the `square' of the Dirac operator, we feel that in an ideal world, in which Lichnerowicz and Peres and everyone else would have been aware of Schr\"odinger's prior contribution, it would have deserved to be called, simply, the `Schr\"odinger formula'.  Indeed not only did Schr\"odinger obtain it 30 years earlier but it is all the more creditable that, as explained in \cite{KayGRG}, he managed to do so without a formula for the spin connection.    In contrast, both Lichnerowicz and Peres benefited from the fact that, by 1963, the notion of spin connection had (as also discussed in \cite{KayGRG}) by then penetrated the realm of the routine in both physics and mathematics.    However, there are other reasons for naming formulae than historical priority.   And perhaps it is good to append some other names so as to distinguish this particular formula of Schr\"odinger from the several other formulae that bear his name.    If so, though, perhaps it should more justly be known as the \emph{Schr\"odinger-Lichnerowicz-Peres formula}.

Let us also take the opportunity of this addendum to correct Footnote \ref{sqrt} which suggested that the square-roots of $g$ in Equation (74) in \cite{Schr32,EngSchr} were unnecessary.   The square roots are in fact needed since, in \cite{Schr32}, and unlike in modern usage, the symbol `$\nabla_k$'  was always taken to mean (\cite[Eq.\ (74)]{Schr32}) $\frac{\partial}{\partial_k} - \Gamma_k$ even when it acts on a spinor vector.

\section{Acknowledgment} 

I thank Gary Gibbons for permission to include the essential content of our addendum to \cite{KayGRG} as Section \ref{Add} above.

\end{document}